\documentclass{aa}
\usepackage{amssymb}
\usepackage{amsmath}
\usepackage{xspace}
\usepackage{graphicx}

\usepackage{natbib}

\def\gr{$\gamma$-ray}

\usepackage{color}

\begin{document}

\title{Strong constraint on hadronic models of blazar activity from Fermi and IceCube stacking analysis}
\author{Andrii Neronov$^1$, Dmitri V. Semikoz$^{2,3,4}$ and  Ksenia Ptitsyna$^5$}
\institute{$^1$Astronomy Department, University of Geneva, Ch. d'Ecogia 16, 1290, Versoix, Switzerland\\
$^2$APC, Universite Paris Diderot, CNRS/IN2P3, CEA/IRFU, Paris, France\\
$^3$Observatoire de Paris, Sorbonne Paris Cite, 119 75205 Paris, France\\
$^4$National Research Nuclear University MEPHI (Moscow Engineering Physics Institute), Kashirskoe highway 31, 115409 Moscow, Russia\\
$^5$Institute for Nuclear Research of the Russian Academy of Sciences,
        60th October Anniversary Prospect 7a, 117312 Moscow, Russia}

\authorrunning{Neronov et al.}
\titlerunning{Constraint on hadronic models of blazars}
\abstract
{
High-energy emission from blazars is produced by electrons which are either accelerated directly (the assumption of leptonic models of blazar activity) or produced in interactions of accelerated protons with matter and radiation fields (the assumption of hadronic models). The hadronic models predict that 
\gr\ emission is accompanied by neutrino emission with comparable energy flux but with a different spectrum.  
}
{
We derive constraints on the hadronic models of activity of blazars  imposed by non-detection of neutrino flux from a population of \gr\ emitting blazars. 
 }
{ 
We stack the \gr\ and muon neutrino flux from 749 blazars situated in the declination strip above $-5^\circ$. 
 }
{
Non-detection of neutrino flux from the stacked blazar sample rules out the proton induced cacade models in which   the high-energy emission is powered by interactions of shock-accelerated proton beam  in the AGN jet with the ambient matter or with the radiation field of the black hole accretion disk. The result remains valid also for the case of interactions in the scattered radiation field in the broad line region.  IceCube constraint could be avoided  if the spectrum of accelerated protons is sharply peaking in the ultra-high-energy cosmic ray range, as in the models of acceleration in the magnetic reconnection regions or in the vacuum gaps of black hole magnetospheres. Models based on these acceleration mechanisms are consistent with the data only if  characteristic energies of accelerated protons are higher than  $10^{19}$~eV.  
}
{}


\maketitle

\section{Introduction}

Supermassive black holes in some 10\% of Active Galactic Nuclei (AGN) accelerate particles and produce jets which could be occasionally aligned along the line of sight \citep{urry}. In this case the AGN appears as a "blazar" detectable in the \gr\ band. The details of the meachanisms of jet generation and particle acceleration in AGN are not clear. High-energy electrons producing the observed  synchrotron and inverse Compton emission could either be accelerated directly (as in leptonic models of AGN activity) or be generated in interactions of high-energy protons  and nuclei (as in hadronic models). 

Different types of hadronic models correspond to different possible mechanisms of acceleration and interactions of high-energy protons and nuclei. The "proton-induced cascade" model(s) ascribe the \gr\ emission to the inverse Compton emission from electromagnetic cascade initiated by the interactions of high-energy protons during their propagation through the radiation field of the AGN central engine and jet \citep{mannheim89,mannheim92,mannheim93}. Otherwise, the \gr\ component of the spectrum could be dominated by the synchrotron radiation from the highest energy protons, as suggested in the "proton synchrotron" models \citep{mucke01,aharonian02,mucke03}. 

The proton induced cascade (PIC) models could be subdivided onto two sub-types depending on the nature of interactions of the high-energy protons.   The cascade could be initiated by interactions of high-energy protons with dense radiation field created by the AGN accretion disk or the jet \citep{mannheim89,begelman90,mannheim92,mannheim93,halzen_zas,neronov02,kalashev14,kalashev15}. Otherwise, the cascade could initiated by interactions of protons with low energy protons from the ambient medium (e.g. from the accretion flow or from the interstellar environment of the black hole) \citep{eichler79,neronov08,ribordy}. 

The details of cascade development depend on the spectrum of primary protons. The most common assumption is that the acceleration site is a relativistic shock in the AGN jet. In this case, the proton spectrum is expected to be a cut-off powerlaw $dN_p/dE\propto E^{-\Gamma}\exp\left(-E/E_{cut}\right)$
with the slope $\Gamma\simeq 2$ and cut-off energy $E_{cut}$ which could reach the Ultra-High-Energy Cosmic Ray (UHECR) energy range. This prediction is somewhat uncertain because of uncertainty of the efficiency of geometry of magnetic field geometry in the AGN jet shocks (turbulence properties, orientation of the ordered field component with respect to the shock normal etc) \citep{ostrowski98,lemoine06,pelletier09}. An alternative possibility for particle acceleration is acceleration in magnetic reconnection regions / events in the AGN accretion flow  \citep{reconnection,reconnection1} or in the jet and acceleration in the vacuum gaps in the black hole magnetosphere \citep{beskin,hirotani98,levinson00,neronov05,neronov07,neronov09,bh_lightning,hirotani16,broderick16,ptitsyna15}. Contrary to the shock acceleration, these types of acceleration mechanisms produce high-energy particle spectra sharply peaked at a characteristic energy, so that the effective $\Gamma$ is $\Gamma\ll 2$.

A straightforward difference between  leptonic and hadronic models of blazar activity is the absence / presence of neutrino emission accompanying the \gr\ emission. No significant neutrino flux is expected in the leptonic models. To the contrary, the specific of hadronic cascades is that the overall neutrino energy flux is always comparable (at least within an order-of-magnitude) with the electromagnetic flux from the source. 

IceCube neutrino telescope has discovered an astrophysical neutrino signal in the energy range in which AGN (and, in particular blazars) are expected to produce neutrino flux \citep{IceCube_1yr,IceCube_PeV,IceCube_3yr,IceCube_2yr,IceCube_combined_2015,IceCube_muonnu_2015,IceCube_6yr}.  Search for neutrino signal from individual blazars or blazar samples did not provide positive detections up to now \citep{tchernin,point_sources}. Also the stacking analysis did not provide solid evidence for neutrino flux from blazars \citep{icecube2014,icecube_stacking,padovani}. Absence of multiplet events in the IceCube muon neutrino data gives additional constraints on the density of neutrino sources which also constrains blazars as possible neutrino sources \citep{waxman}.  However, time coincidence of a neutrino arrival with a blazar outburst was claimed to support the hypothesis of blazar origin of astrophysical neutrino signal  \citep{kadler16}. 

Non-detection of the signal from the \gr\ brightest blazars was used by \citet{tchernin} to derive limits on the parameters of the PIC hadronic models of blazars.  The data of IC-40 detector limit $E_{cut}$ and $\Gamma$ to be roughly 
\begin{equation}
E_{cut}\gtrsim 10^{18}\mbox{ eV,}\ \  \ \Gamma\lesssim 2
\end{equation}

In what follows we combine the updated IceCube neutrino data \citep{IceCube_6yr} with the data of  Fermi \gr\ telescope to  improve previously derived bounds on $\Gamma$ and $E_{cut}$. We use stacking, rather than individual source analysis and  consider cumulative  \gr\ and neutrino spectra of a large number of Northern sky blazars.  

Our results severely limit the class of PIC models consistent with the the combined \gr\ and neutrino data. Most of the models based on assumption of shock acceleration mechanisms appear to be ruled out. Models which avoids the \gr\ $+$ neutrino constraint are those predicting the accelerated particle spectra sharply peaked in the UHECR range. However, also in these models the parameters of particle acceleration mechanisms have to be tuned to provide the highest possible efficiencies. 

\section{Stacking analysis approach to \gr\ and neutrino spectra}

Within the PIC models, the overall energy flux of neutrinos and photons from the source is comparable, at least within order-of-magnitude
\begin{equation} 
\label{eq:gamma_nu}
F_\nu\sim {F_\gamma}
\end{equation}
The neutrino signal from individual blazars is detectable in backgorund-free regime in the energy band above several hundred TeV \citep{IceCube_6yr}.  Suppose that the upper limit on neutrino flux from an individual source is $F_{\nu,lim}$. Non-detection of neutrinos from any of $N$ stacked sources provides an upper limit on "typical" neutrino flux from a source
\begin{equation}
\label{eq:cum_nu}
{\cal F}_{\nu}\lesssim \frac{F_{\nu,lim}}{N}
\end{equation}
The upper limits on the flux of individual sources in the Northern hemisphere reported by IceCube \citep{point_sources} range between $F_{\nu,lim}\simeq 10^{-9}$~GeV/(cm$^2$s) and $3\times 10^{-9}$~GeV/(cm$^2$s), for the declinations ranging from 0 to 90 degrees. The best sensitivity is achieved in the energy range around  100~TeV, where the atmospheric neutrino background becomes low enough, so that the signal with the flux level $F_{\nu,lim}$  is detectable in nearly background free regime.   The limiting flux level for individual sources is comparable to the typical flux level of of blazars detected by Fermi/LAT telescope. Thus, source by source analysis could only mildly constrain the limits of the PIC type models by imposing a requirement that the bulk of the neutrino flux is not emitted in the 100~TeV band. This limits the properties of the parent proton spectrum, as reported by \citet{tchernin}.   

A size of the population of GeV \gr\ detected blazars is $\sim 10^3$. Although the flux of most of the Fermi/LAT detected blazars is typically one or two orders of magnitude lower than the fluxes of the brightest blazars, the cumulative \gr\ flux of all the detected blazars is still an oder-of-magnitude larger than that of an individual bright blazar. Thus, the expected cumulative neutrino signal from the population of the \gr\ detected blazars is a factor of 10 larger than the flux from any individual bright blazar. If the typical energies of neutrinos from blazars are much higher than 100~TeV (as suggested by the IceCube constraints derived from the analysis of individual bright blazars \citep{tchernin}, the signal from individual blazars is not detectable, but an oder-of-magnitude stronger cumulative neutrino signal from \gr\ loud blazar population might still be detectable at the highest energies.  

\section{Data analysis}

\subsection{Fermi/LAT}

\begin{figure*}
\includegraphics[width=\linewidth]{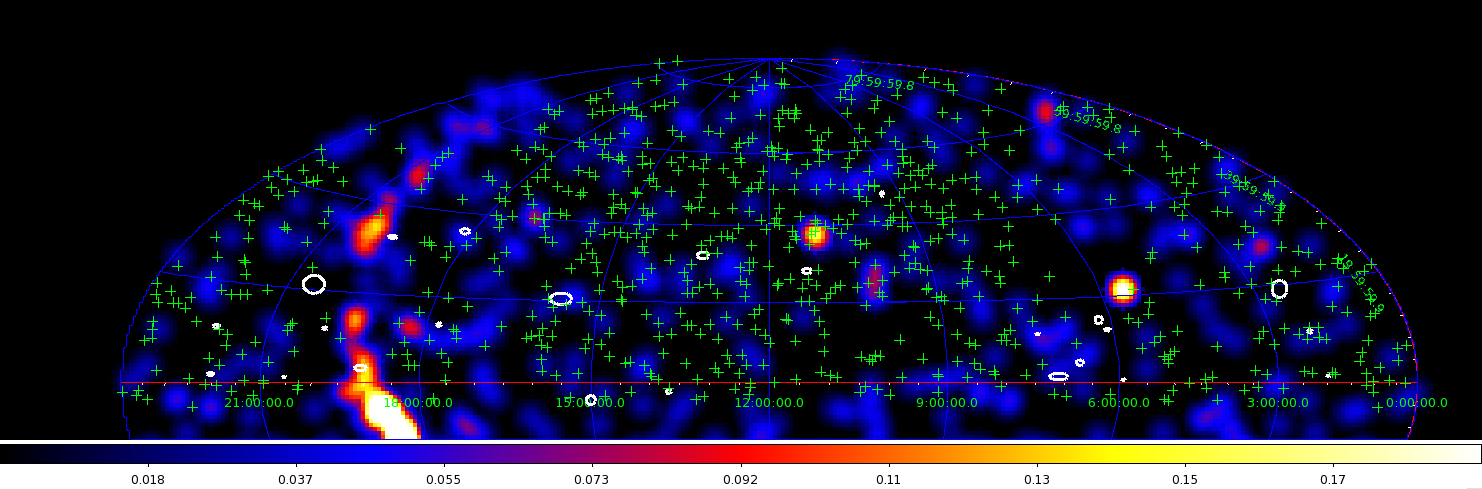}
\caption{IceCube muon neutrino events (white ellipses) with direction uncertainty less than 4 degrees, overlaid over Fermi/LAT countmap pf the Northern sky in the energy range above 1~TeV, smoothed with 3 degree Gaussian. Green crosses show blazars selected for the stacking analysis.}
\label{fig:image}
\end{figure*}

For our stacking analysis we consider publicly available data of Fermi/LAT telescope collected during the time period between August 2008 and June 2016. We filter the LAT event list with the help of {\it gtselect} -- {\it gtmktime} tools following the recommendations of the Fermi/LAT team\footnote{http://fermi.gsfc.nasa.gov/ssc/data/analysis/} to select only events belonging to the {\tt CLEAN} sub-class. For each selected source we extract the spectrum using the aperture photometry method estimating the exposure with the help of {\it gtexposure} tool. We sum the source counts, the  background counts and the exposures in the direction of all the selected sources to produce a cumulative source spectrum. For each source, the source signal is collected from a circle of the radius $1^\circ$ around the source position. The background is estimated from a $1^\circ$ circle displaced by $2^\circ$ from the source position. 

The list of blazars selected for the stacking analysis includes $N=749$  Flat Spectrum Radio Quasars (FSRQ) and BL Lacerta (BL Lac) type objects at declinations $DEC>-5^\circ$    listed in the third Fermi/LAT source catalog \citep{fermi_cat}. This choice of the declination range is determined by sky region from which the IceCube astrophysical neutrino signal  is collected \citep{IceCube_6yr}. 

Fig. \ref{fig:spectrum} shows the resulting cumulative spectrum of blazars in this part of the sky. One could see that it is nearly identical to the cumulative spectrum of high Galactic latitude sources found in the Ref. \cite{fermi_igrb}. This is not surprising, given the fact that blazars constitute the dominant extragalactic source population.

\subsection{IceCube}

\begin{figure}
\includegraphics[width=\linewidth]{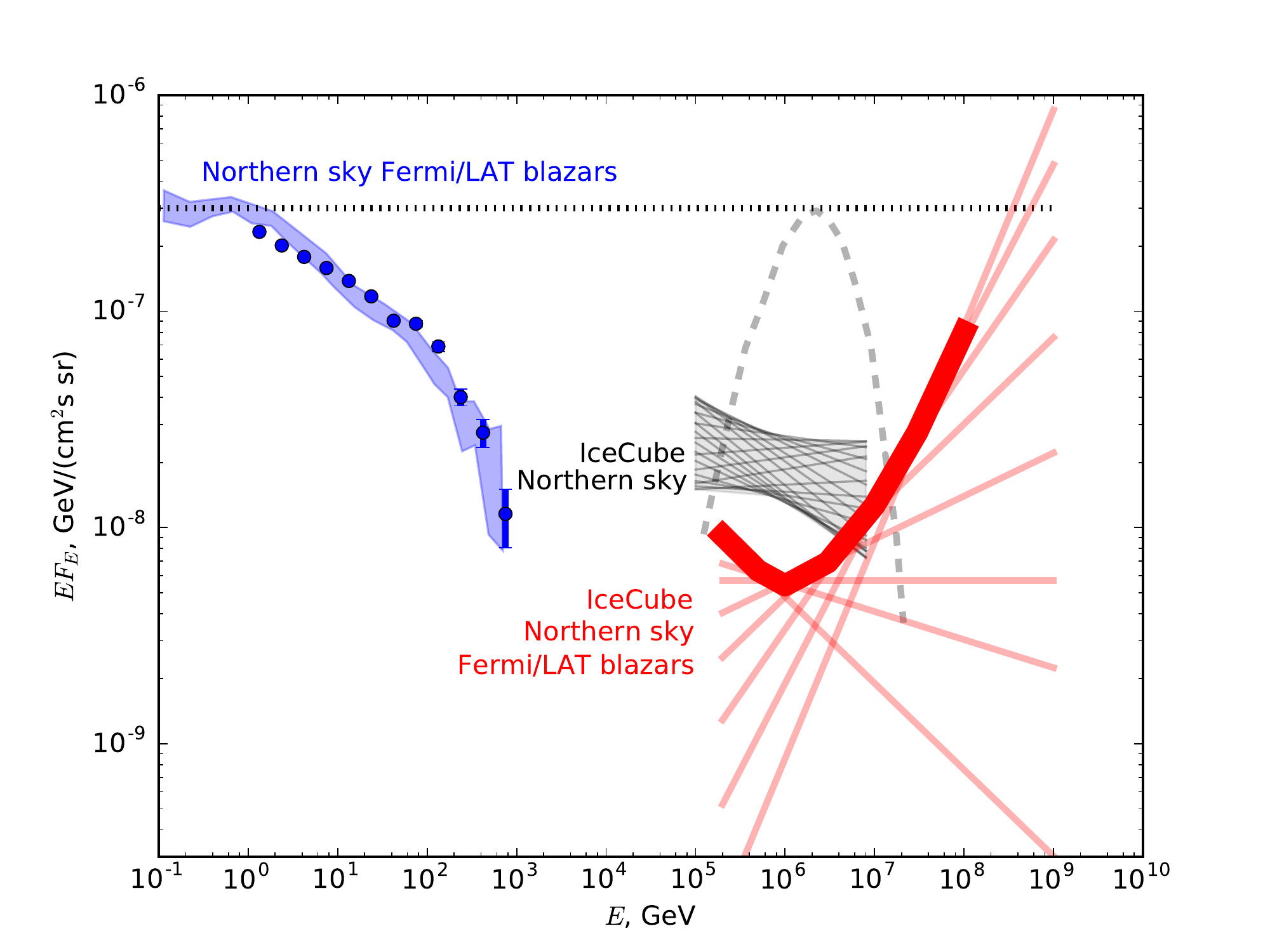}
\caption{Cumulative \gr\ (blue data points) spectrum and neutrino flux upper limit (red) for the Northern hemisphere blazars. Blue shaded band shows the spectrum of extragalactic sources resolved by Fermi telescope, from \citet{fermi_igrb}. Black hatched bow-tie shows the IceCube astrophysical neutrino flux in the muon neutrino channel from \citet{IceCube_6yr}. Grey dashed line shows a model neutrino spectrum for a PIC model from \citet{tchernin}.}
\label{fig:spectrum}
\end{figure}

We use the list of 29 muon neutrino events with energy proxies above 200~GeV reported by \citet{IceCube_6yr}. Three out of the 29 events have large statistical uncertainty of direction reconstruction (larger than $3^\circ$). The chance coincidence probability for such events to have one of the 749 blazars within their error ellipse is of the order of one. We remove these events from the analysis. Rejection of these events reduces the effective exposure of IceCube data set by $3/29\simeq 10\%$. 

The $90\%$ error ellipses of the 26 neutrino events retained for the analysis do not contain \gr\ detected  blazars, except for one blazar, OP 313, which is at the border of the error ellipse of the muon neutrino event 36. This is consistent with a chance coincidence expectation. The event number 36 has the energy proxy $E=200$~TeV and only 0.45 "signalness" value \citep{IceCube_6yr}, i.e. is is more likely to be part of the atmospheric neutrino background. A 90\% upper limit on the number of muon neutrino events with energy proxies above 200 TeV from blazars is 
\begin{equation}
N_{lim}=4.
\end{equation}

The spectrum of neutrino emission from blazars is, in general, unknown. To derive an upper limit on the neutrino flux from blazars for arbitrary spectral shape, we calculate the maximal possible normalization  $\kappa$ of a powerlaw neutrino flux
\begin{equation}
\frac{dN_\nu}{dE}=\kappa \left(\frac{E}{E_*}\right)^{-\Gamma}
\end{equation}
(where $E_*$ is normalization energy fixed here to $E_*=1$~PeV) for different slopes $\Gamma$. We scan over the slopes $\Gamma$ to find an "envelope" curve of the different maximal possible flux powerlaws, as described by \cite{tchernin}. Realistic neutrino emission spectra could typically be well approximated by powerlaws in the energy range where the IceCube sensitivity is highest (about PeV for the considered IceCube data set), unless the spectrum has a high or low energy cut-off exactly in the IceCube sensitivity range. This implies that  envelope curve of the upper limits on the powerlaw type spectra could be also used for the realistic spectra: the spectra consistent with the data could at most "touch" the envelope curve from below. 

To derive the envelope curve, we note that the IceCube exposure for $\nu_\mu$ (or ${\overline \nu}_\mu$) in the energy range above 400 TeV is well approximated by a powerlaw
\begin{equation}
T_{exp}A_{eff}\simeq TA_*(E_\nu/E_*)^{p}
\end{equation}
with the normalisation $TA_*\simeq (7/2)\times 10^{14}$~cm$^2$s (averaged over the solid angle $\Omega=2\pi(1-\cos(95^\circ))$ and counting only muon neutrinos) at a reference energy $E_*=1$~PeV and the slope $p=0.34$  \citep{IceCube_6yr}.   

The expected number of muon neutrinos  and anti-neutrinos in an energy range $E_{min}<E_\nu<E_{max}$ for a given flux normalization $\kappa$ is
\begin{eqnarray}
\label{nmu}
&&N_{\nu_\mu}=\frac{1}{3}\kappa\int \Omega T_{exp}A_{eff}(E_\nu)\left(\frac{E_\nu}{E_*}\right)^{-\Gamma}dE_\nu= \nonumber\\
&&\frac{\kappa \Omega TA_*E_*}{3(p-\Gamma+1)}\left(\left[\frac{E_{max}}{E_*}\right]^{p-\Gamma+1}-\left[\frac{E_{min}}{E_*}\right]^{p-\Gamma+1}\right)
\end{eqnarray} 
where the factor $1/3$ accounts for the fact that the muon neutrinos constitute $1/3$ of the signal (adopting standard assumptions about production mechanism and mixing). 

Muons produced in the charged current interactions outside the IceCube detector have initial energies $E_\mu\simeq (1-y_{cc})E_\nu$ where $y_{cc}$ is the average inelasticity of the charged current interactions \citep{gandhi}. The energies of most detected neutrino-induced muons  are  much lower than the initial muon energy because of the energy loss in the rock $dE_\mu/dx=-(a+bE_\mu)$  \citep{chirkin}, so that  the distribution of final energies of muons originating from monoenrgetic neutrinos with energy $E_\nu$ steadily undergoing charged current interactions all over a large distance through the rock / ice is $dN_\mu/dE\propto E^{-1}$ in the energy range $E<E_0$, down to the energy $E_{crit}=a/b\simeq 1$~TeV. This suggests the probability density function for the muon energy \citep{ribordy1}
\begin{equation}
\frac{dp(E_\mu,E_\nu)}{dE}=\left(\ln\left(1+\frac{(1-y_{cc})E_{\nu}}{E_{crit}}\right)\right)^{-1}\frac{1}{(E_\mu+E_{crit})}
\end{equation} 
The distribution of the energies of the muon events is then
\begin{eqnarray}
&&\frac{dN_\mu}{dE_\mu}=\int_{\frac{E_\mu}{(1+y_{cc}}}^\infty\frac{dp(E_\mu,E_\nu)}{dE}T_{exp}A_{eff}(E_\nu)\Omega\frac{\kappa}{3}\left(\frac{E_\nu}{E_*}\right)^{-\Gamma}dE_\nu\nonumber\\
&&\simeq\frac{\Omega TA*}{3\ln\left(\frac{E_\mu}{E_{crit}}\right)(p-\Gamma+1)}\left(\frac{E_\mu}{E_*}\right)^{p-\Gamma}
\end{eqnarray}
Integrating the muon distribution in the energy range $(E_{min},E_{max})$ one finds
\begin{eqnarray}
N_{\mu}&\simeq &\frac{\kappa \Omega TA_*E_*}{3(\Gamma-p-1)}\times\\
&&\left(\left[\frac{(E_{min}/E_*)^{p-\Gamma+1}}{\ln\left(E_{min}/E_{crit}\right)}\right]-\left[\frac{(E_{max}/E_*)^{p-\Gamma+1}}{\ln\left(E_{max}/E_{crit}\right)}\right]\right)\nonumber
\end{eqnarray} 
which differs from the neutrino number by a factor $\left(\ln\left(E_\mu/E_{crit}\right)\right)^{-1}\simeq 0.2$.

Normalizing $N_\mu=N_{lim}$  one finds $\kappa$ for different values of $\Gamma$.  The result is shown by the red straight lines in Fig. \ref{fig:spectrum}. The red thick curve shows the envelope of all maximal allowed powerlaws. It is useful to note that in $\Gamma<p+1\simeq 1.3$, the signal statistics is dominated by the highest energy events and the maximal normalization of the powerlaw depends on the high-energy cut-off in the spectrum. We have fixed $E_{max}$ to $10^{18}$~eV in our calculation.

\section{Results and discussion}

\subsection{Inconsistency of the data with proton induced cascade model with shock-accelerated protons in the jet}

One could see from Fig. \ref{fig:spectrum} that the upper limit on the neutrino flux from blazars in the PeV energy range  is two orders of magnitude below the expected flux level.
The mismatch between the \gr\ flux level and neutrino flux upper limits rule our hadronic models in which neutrino spectra are expected to produce most of the power in the PeV energy range.  

The pion production in $p\gamma$ interactions occurs only above an energy threshold
\begin{equation}
E_{p,thr}=10^{16}\left[\frac{\epsilon}{10\mbox{ eV}}\right]^{-1}
\end{equation}
where $\epsilon$ is the characteristic energy of photons from the radiative environment of the AGN. The energies of neutrinos originating from the charged pion decays are typically below $\lesssim 10\%$ of the parent proton energy: 
\begin{equation}
\label{eq:enu}
E_\nu\sim 10^{15}\left[\frac{\epsilon}{10\mbox{ eV}}\right]^{-1}
\end{equation}
Pion decays transfer nearly equal power to the neutrino and electromagnetic emission. Development of electromagnetic cascade in the source transfers the electromagnetic power to the GeV-TeV energy band. Horizontal dashed line in Fig. \ref{fig:spectrum} shows an estimate of the electromagnetic power from the blazar population. Contrary to the electromagnetic power, the neutrino emission power stays in the energy band (\ref{eq:enu}). 

This is the case for the proton-induced cascade models in which the shock-accelerated protons interact with the radiation field of accretion disk. Conventional geometrically-thin / optically-thick accretion disks in AGN have temperatures reaching $10^4$~K in the innermost portions of the disk close to the last stable orbit \citep{shakura,accretion_power}. Protons accelerated near the black hole or in the innermost portion of the AGN jet interact with the direct UV radiation form the disk or with the disk radiation scattered in the Broad Line Region. This inevitably produces neutrinos with energies in the PeV range. An example of neutrino spectrum calculated for proton spectrum with $\Gamma=2$ and $E_{cut}=10^{17}$~eV interacting with  the soft photons from accretion disk with the spectrum peaking at $\epsilon=15$~eV  (from \citet{tchernin}) is shown by the grey dashed line in Fig. \ref{fig:spectrum}.  

\begin{figure}
\includegraphics[width=\linewidth]{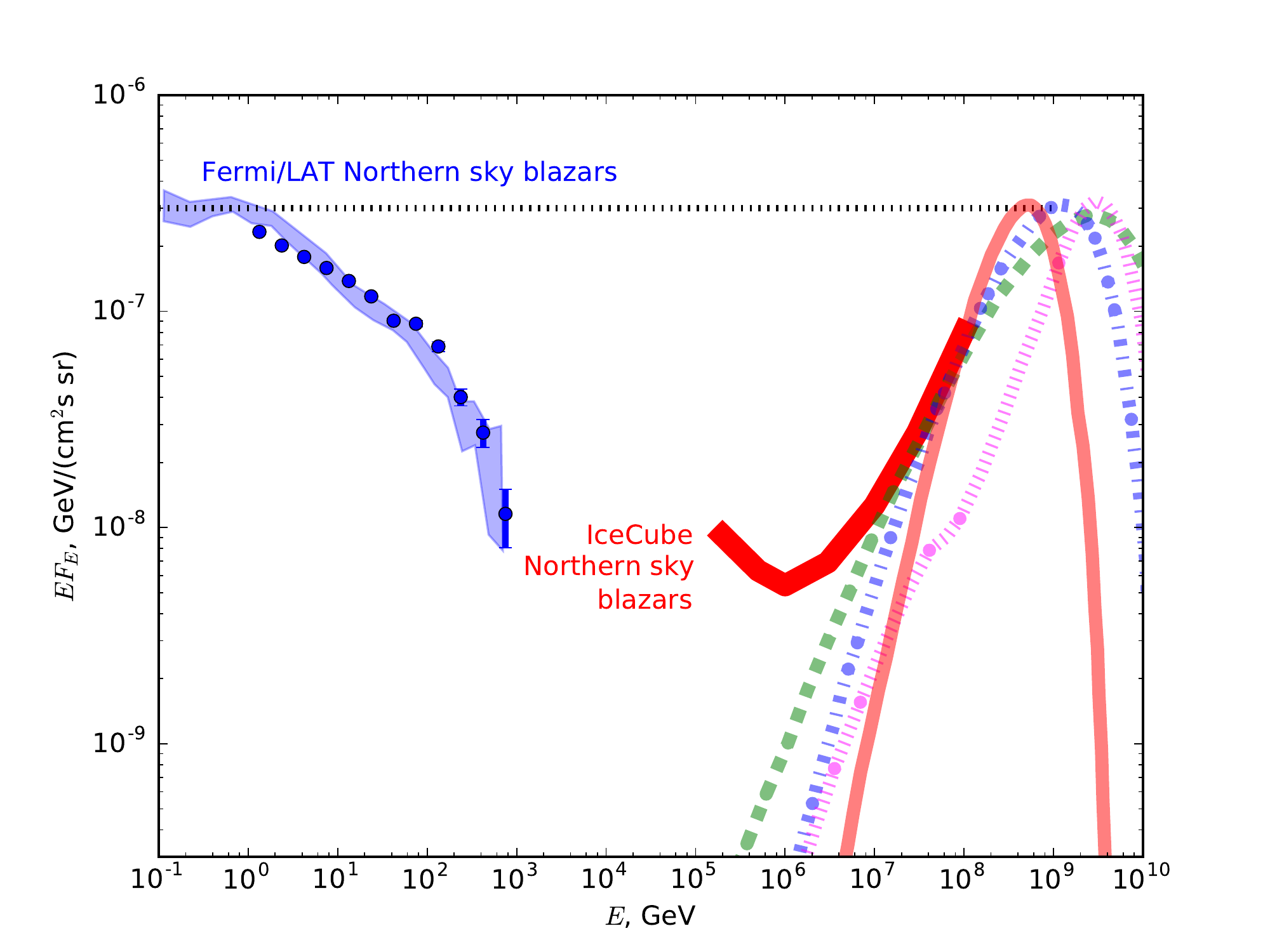}
\caption{Comparison of the IceCube upper limit on the neutrino flux with predictions of PIC models of blazars. }
\label{fig:spectrum1}
\end{figure}

The IceCube bound could be avoided if the bulk of neutrino power is emitted in an energy band different from 0.1-10~PeV. This is the case if the soft photon targets for $p\gamma$ interactions are the in the infrared or microwave range. This type of interactions is considered in the model of \citet{kusenko,kusenko1}. The characteristic energy of the Cosmic Microwave Background photons is $\epsilon\simeq 10^{-3}$~eV. Magenta dotted line in Fig. \ref{fig:spectrum1} shows a representative model of neutrino spectrum  from interactions of protons with the spectrum $E^{-2}$ with high-energy cut-off at $10^{20}$~eV during their propagation through the CMB and Extragalactic Background Light radiation fields \citep{kusenko1}. Normalizing the model flux on the average \gr\ flux from blazars one finds that the model spectrum is consistent with the IceCube constraint. 

Another type of hadronic models severely constrained by combined IceCube $+$ Fermi data are the models in which shock accelerated protons interact with the ambient medium protons and nuclei. In this case the pion production threshold is in the 100~MeV range and the neutrino spectrum in the energy range above the threshold approximately repeats the proton spectrum. If the proton spectrum is a powerlaw with the slope close to 2, the neutrino spectrum  in the PeV energy range is also a powerlaw with the slope close to 2 and with the energy flux comparable to that of the \gr\ flux, within an order-of-magnitude. The mismatch between the maximal possible normalization of an $E^{-2}$ type neutrino spectrum and the dashed horizontal line in Figs. \ref{fig:spectrum}, \ref{fig:spectrum1} is two orders of magnitude, which means that the model is ruled out. 

\subsection{Constraints on models with sharply peaked proton spectra}

The  IceCube $+$ Fermi constraint on the proton-induced cascade model could be avoided if the high-energy protons are not produced by the shock acceleration process. If the protons are injected by acceleration taking place in large scale electric fields, as it is the case for the field in magnetic reconnection regions or in the vacuum gaps in the black hole magnetosphere, the spectrum of protons is sharply peaked at a particular energy, rather than has a powerlaw shape. If the characteristic energy of the acceleration process is large enough, the peak energy of the neutrino spectrum is determined by the characteristic proton energy, rather than by the threshold of the $p\gamma$ reaction. Blue dash-dotted line in Fig. \ref{fig:spectrum1} shows the neutrino spectrum form interactions of protons with energies $3\times 10^{19}$~GeV interacting with $\epsilon=10$~eV photons.  The low energy part of the model spectrum is tangent to the envelope of the IceCube upper limits. This means that the models of PIC in the AGN accretion disk radiation fields are consistent with the limits if the proton spectrum has a sharp low energy cut-off 
\begin{equation}
 E_p\gtrsim 2\times 10^{19}\mbox{ eV, (PIC in UV radiation field)} 
\end{equation}

An alternative possibility for avoiding the IceCube$+$ Fermi constraint is to consider  models in which the accretion on the black hole forms a radiatively inefficient accretion flow (RIAF). This type of models are believed to be applicable to low luminosity radio galaxies and BL Lac type objects. In this case the low energy radiation from the accretion flow is the synchrotron radiation from electrons heated to relativistic temperatures by collisions with protons. The synchrotron radiation of RIAF peaks in the infrared range \cite{riaf}, $\epsilon\lesssim 0.1$~eV, and the neutrino spectrum peaks in the energy range above $10^{17}$~eV. This is illustrated by the red solid curve in Fig. \ref{fig:spectrum1} which is the neutrino spectrum produced in interactions of $10^{19}$~eV protons with $0.1$~eV photons. One could see that the model spectrum is consistent with the IceCube data, so that models with proton spectrum with low-energy cut-off at
\begin{equation}
 E_p\gtrsim 0.6\times 10^{19}\mbox{ eV, (PIC in IR radiation field)} 
\end{equation}
are also consistent with the data. 

The IceCube $+$ Fermi constraint could be also avoided in the PIC model where high-energy protons interact with low energy protons from the accretion flow, if the high-energy proton spectrum is sharply peaked at high energies.  Green dashed line in Fig. \ref{fig:spectrum1} shows the calculation of the model neutrino spectrum from interactions of protons with energy $2\times 10^{20}$~eV with the low energy protons. One could see that the low energy part of the neutrino spectrum is tangent to the IceCube upper bound envelope curve. This means that models based on $pp$ interactions are constrained to produce extremely high-energy protons: 
\begin{equation}
 E_p\gtrsim 1\times 10^{20}\mbox{ eV (PIC initiated by}\ pp\mbox{ interactions)}
\end{equation}

\subsection{Proton energies in black hole magnetospheric gap models}

\begin{figure*}
\includegraphics[width=0.8\linewidth]{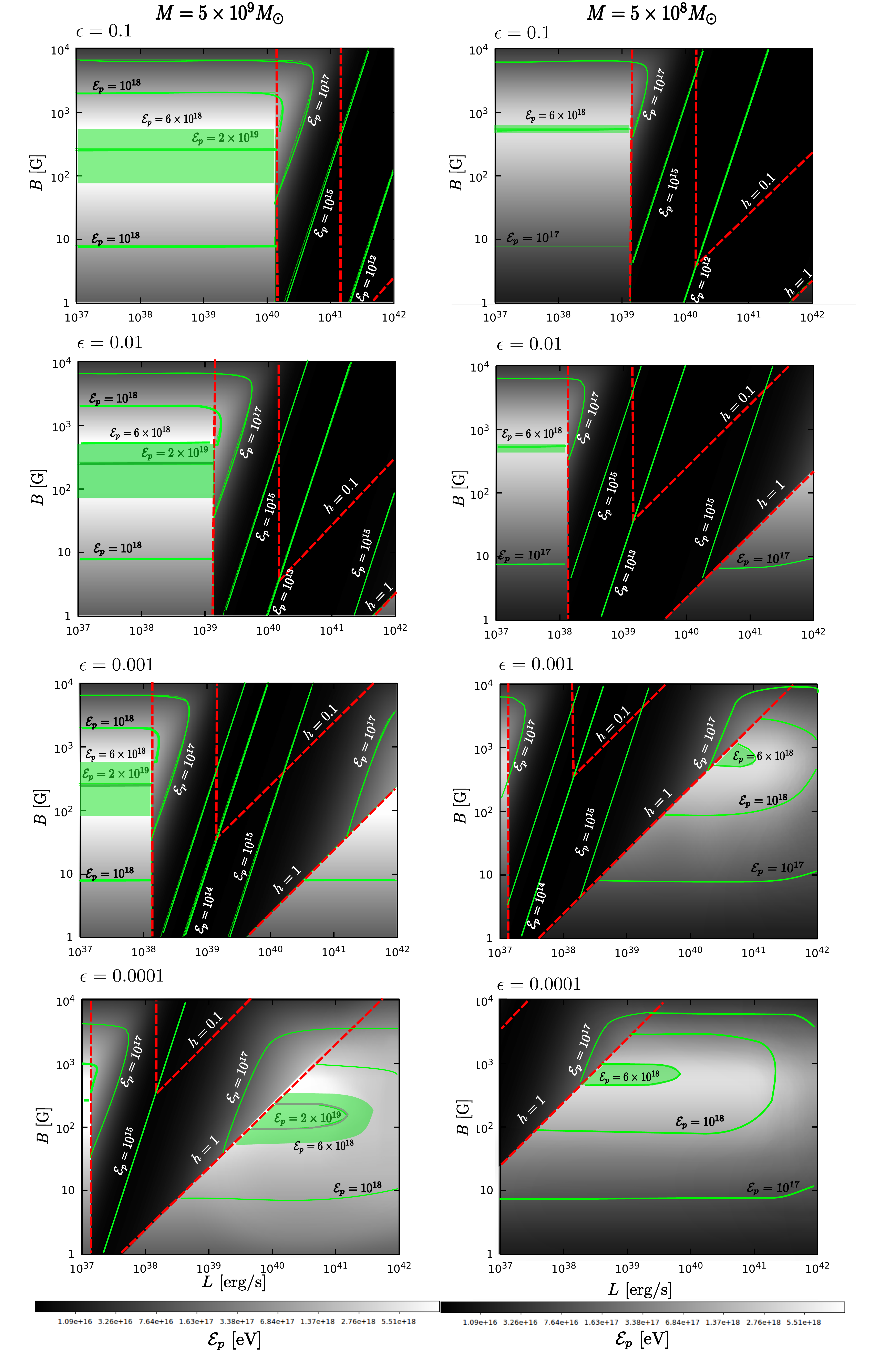}
\caption{Energies of protons accelerated in magnetospheric vacuum gaps near a black hole of mass $M=3\times 10^9M_\odot$ (left column) or $3\times 10^8M_\odot$ (right column) surrounded by RIAF with synchrotron emission spectrum peaking in the infrared at $\epsilon=0.1$~eV (top row) to $10^{-4}$ eV (bottom row). Red dashed  lines show the dimensionless gap height $h$. Green solid curves show proton energies. Dark/grey parts of the diagram show the parameter range expluded by the IceCube+Fermi data set. Green shaded areas correspond to the allowed range of parameters.   }
\label{fig:Ep}
\end{figure*}

Energies of protons accelerated in the gap of the height $H$ in the magnetosphere of a black hole of the mass $M$ with the height  are limited by the finite extent of the gap which is defined by the onset of electron-positron pair production on the soft photon background field present in the magnetosphere. The gap height depends on the luminosity $L$ and size $R$ of the soft photon field as well as on the characteristic soft photon energy $\epsilon$ as well as on the rate of electron/proton acceleration in the gap, which is determined by the specific angular momentum of the black hole $a$ \citep{beskin,hirotani98,levinson00,neronov05,neronov07,neronov09,bh_lightning,hirotani16,broderick16,ptitsyna15}. .

Fig. \ref{fig:Ep} shows the attainable proton energy   as a function of source luminosity $L$ and magnetic field $B$, calculated within the framework discussed by \citet{ptitsyna15}. The assumption of the model is that the black hole accretes in the RIAF mode in which the soft photon field is produced via synchrotron emission from electrons heated to the temperatures 10-100~MeV by the protons. The RIAF synchrotron radiation spectrum typically peaks in the infrared range (as opposed to the UV dominated spectrum of optically thick geometrically thin accretion disk). Different columns of he figure correspond to two different black hole masses. Different rows correspond to different soft photon fields in the AGN central engine. In all the cases the infrared / microwave soft photon source is supposed to be distributed over a region of the size $10 R_{Schw}$, where $R_{Schw}$ is the Schwarzschild radius of the black hole. 

The process of the pair production starts to limit the gap height when the luminosity reaches certain (magnetic field dependent) value around $10^{40}$~erg/s. At lower luminosities of the RIAF the density of the soft photon field is not sufficient for the pair production within the extent of the black hole magnetosphere on the distance scale $R\sim R_{Schw}$ about the Schwarzschild radius of the black hole. In this case the limiting proton energy  is estimated as 
\begin{equation}
E_p\sim eBR_{Schw}\simeq 10^{19}\left[\frac{B}{100\mbox{ G}}\right]\left[\frac{M}{3\times 10^9M_\odot}\right]\mbox{ eV}
\end{equation}
where $M$ is the black hole mass. The limit $E_p>6\times 10^{18}$~eV converts into a lower bound on magnetic field strength in the innermost part of the RIAF:
\begin{equation}
B>100\left[\frac{M}{3\times 10^9M_\odot}\right]^{-1}\mbox{ G}
\end{equation}

Proton energies drop below $6\times 10^{18}$ eV also at high values of $B$. In this case the proton synchrotron loss limits the energies of protons, as in the models of Ref. \citet{mucke01,aharonian02,mucke03}.  This type of models is not directly constrained by a combination of the Fermi/LAT and IceCube data discussed above. 

 The pair production initiated by electrons accelerated in the gap also gets suppressed at high luminosities because the strong inverse Compton loss rate does not allow electrons to get accelerated to the energies needed for the pair production. Suppression of the pair production at low and high luminosities allows acceleration of protons to the energies in excess of $10^{19}$~eV. However, in this case protons start to produce pairs themselves, when accelerated to the energies higher than the pair production threshold. This limits the proton energies also in the case of high luminosity RIAF as it is clear from Fig. \ref{fig:Ep}.

The  green shaded regions in different panels of Fig. \ref{fig:Ep} show the ranges of $L,B$ parameter space in which proton energies reach $>6\times 10^{18}$~GeV. Acceleration in the magnetospheric vacuum gaps near black holes surrounded by RIAF with such parameters would produce neutrino and electromagnetic emission consistent with IceCube and Fermi/LAT data. One could see that the IceCube plus Fermi/LAT constraints could be satisfied in only very limited range of parameter space.

\section{Conclusions}

We have shown that a combination of IceCube and Fermi/LAT data rules out certain types of hadronic models of emission from blazars. The models inconsistent with the data are those in which the observed \gr\ emission is produced by a particle cascade initiated by shock accelerated protons in the UV radiaiton field of the AGN central engine. 

Hadronic models consistent with the data are those in which high-energy protons spectra are sharply peaked in the UHECR range. An example of this type of models is the model of proton acceleration in the vacuum gaps of black hole magnetospheres. We have shown that the IceCube and Fermi/LAT data constrain the parameter space of such models (luminosity and magnetic field in the RIAF surrounding the black hole). Models consistent with the data predict neutrino flux originating from UHECR production in the blazars. This suggests that the model is testable via observations of neutrino flux in the energy range higher than that accessible with IceCube, $E_\nu\sim 0.1-1$~EeV. Increase of IceCube exposure, or exploration of this energy range with dedicated detectors optimized for the 0.1-1~EeV range, like CHANT \citep{chant}, ARA \citep{ara} could be used to test the model. 

\section*{Acknowledgements}

We would like to thank S.Shoenen for useful discussions on details of IceCube results. The work of KP is supported by the Russian Science Foundation grant 14-12-01340.

\bibliography{No_proton_blazar}

\end{document}